\documentclass{aa}  
\usepackage{graphicx}
\usepackage{natbib}
\usepackage{url}
\usepackage{txfonts}
\begin{document}
   \title{Evolution of sunspot properties during solar cycle 23}

%   \subtitle{blah}

   \author{F. T. Watson\inst{1} \and L. Fletcher \inst{1} \and S. Marshall
\inst{2}}

   \offprints{F. T. Watson}

   \institute{School of Physics and Astronomy, SUPA, University of Glasgow,
Glasgow G12 8QQ, U.K.
              \email{f.watson@astro.gla.ac.uk, lyndsay.fletcher@glasgow.ac.uk}
              \and
              Department of Electronic and Electrical Engineering, University of
Strathclyde, Glasgow G1 1XW, U.K.
              \email{s.marshall@eee.strath.ac.uk}
              }

   \date{Received ; accepted }

% \abstract{}{}{}{}{} 
% 5 {} token are mandatory
 
  \abstract
  % context heading (optional)
  % {} leave it empty if necessary  
   {The long term study of the Sun is necessary if we are to determine the
evolution of sunspot properties and thereby inform modeling of the solar dynamo,
particularly on scales of a solar cycle.}
  % aims heading (mandatory)
   {We aim to determine a number of sunspot properties over cycle 23 using the
uniform database provided by the SOHO Michelson Doppler Imager data. We focus in
particular on their distribution on the solar disk, maximum magnetic field and
umbral/penumbral areas. We investigate whether the secular decrease in sunspot
maximum magnetic field reported in Kitt Peak data is present also in MDI data.}
  % methods heading (mandatory)
   {We have used the Sunspot Tracking And Recognition Algorithm (STARA) to
detect all sunspots present in the SOHO Michelson Doppler Imager continuum data
giving us 30\,084 separate detections. We record information on the sunspot
locations, area and magnetic field properties as well as corresponding
information for the umbral areas detected within the sunspots, and track them
through their evolution.}
  % results heading (mandatory)
   {We find that the total visible umbral area is 20-40\% of the total visible
sunspot area regardless of the stage of the solar cycle. We also find that the
number of sunspots observed follows the Solar Influences Data Centre
International Sunspot Number with some interesting deviations. Finally, we use
the magnetic information in our catalogue to study the long term variation of
magnetic field strength within sunspot umbrae and find that it increases and
decreases along with the sunspot number. However, if we were to assume a secular
decrease as was reported in the Kitt Peak data and take into account sunspots throughout the whole solar cycle we would find the maximum umbral
magnetic fields to be decreasing by 23.6 $\pm$ 3.9 Gauss per year, which is far
less than has previously been observed by other studies (although measurements
are only available for solar cycle 23). If we only look at the declining phase of cycle 23 we find the decrease in sunspot magnetic fields to be 70 Gauss per year.}
  % conclusions heading (optional), leave it empty if necessary 
   {}


   \maketitle
%
%________________________________________________________________

\section{Introduction}
Sunspots are dark areas on the solar surface and are associated with strong
magnetic fields. The magnetic field inhibits the convective flow of plasma in
the region and as this is the primary mechanism for heat transport at the
surface, the sunspot is cooler and darker. Study of sunspots started around the
early 1600s although there are records of observations in China going back for
2000 years \citep{Yau1988,Eddy1989}. Since the discovery of the magnetic field
in sunspots \citep{Hale1908} they have been a primary indicator of solar
activity and detailed records have been kept. By studying the evolution of
sunspot characteristics (area, field strength, etc), on timescales of days  we
can gain insight into their formation and dispersal, while studies on longer
timescales (months and years) can reveal the longer-term behaviour of the Sun's
large-scale magnetic field, naturally of great importance for constraining
models of the solar dynamo.  For example, the North-South asymmetry of sunspot
numbers and areas is well-established and has been studied for many decades
\citep[see e.g. ][ and references
therein]{1993A&A...274..497C,2006AdSpR..38..868Z,2007A&A...476..951C} and may
indicate a phase lag between the magnetic activity in the northern and southern
hemispheres, possibly hinting at non-linear behaviour, such as random
fluctuations of the dynamo terms and strong high order terms
\citep[e.g.][]{2003A&ARv..11..287O}.

The sunspot cycle variation of many solar parameters is of course well
established, however it was reported by \citet{Penn2006} that Zeeman
splitting observations of the strongest fields in sunspot umbrae show a secular
decrease between 1998 and 2005, apparently without a clear cyclic variation.
This goes hand-in-hand with an increase in the umbral brightness. Such a secular
change, if verified, would have striking implications for the coming sunspot
cycles - \citet{2010arXiv1009.0784P} suggest that if the trend continues there
would be virtually no sunspots at the time of cycle 25. It is one of the main
goals of the present study to automatically examine the MDI data for such
behaviour. In creating the dataset necessary to do this we also obtain and
report on the cycle-dependent behaviour of sunspot areas and locations. In
particular, the total projected area of sunspots present on the visible disk is
of interest in solar spectral irradiance studies
\citep{1982JGR....87.4319W,1985SoPh...97...21P,1997SoPh..173..427F} where it
enters as a parameter in spectral irradiance calculations. 

We are fortunate now to have long and consistent series of solar observations
from which such parameters can be extracted, and the computational capacity to
do it automatically. Image processing and feature recognition/tracking in solar
data is now a very active field \citep{2010SoPh..262..235A}, and sunspot
detection is a well-defined image processing problem that has been studied by
several authors
\citep{2005SoPh..228..377Z,2008SoPh..248..277C,2008SoPh..250..411C,Watson2009}.
It is the purpose of this article to  detail some physical properties of
sunspots detected in the continuum images from the SOHO/MDI instrument
\citep{Scherrer1995} and how they vary throughout solar cycle 23. We have used
an image processing algorithm based on mathematical morphology
\citep{Watson2009}.

The article proceeds with section 2 detailing the generation of the sunspot
catalogue and the results of looking at evolution in sunspot area and locations
over solar cycle 23. Then, section 3 details the evolution of magnetic fields in
sunspots, particularly in the umbra where the fields are strongest. Finally, in
section 4 we finish with our discussion and conclusions.

\begin{figure}[ht]
\centerline{\includegraphics[width=0.45\textwidth,clip=]{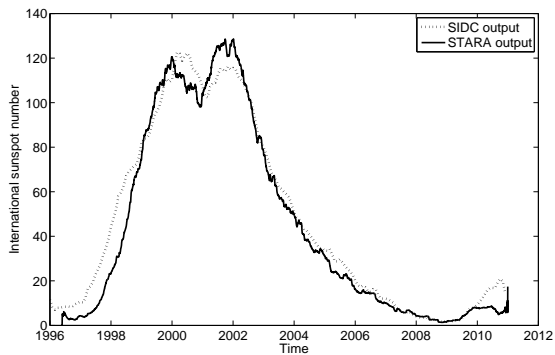}}
\caption{The solid line shows the number of sunspots detected by the STARA code,
scaled to match the magnitude of the international sunspot number near the peaks
as calculated by the SIDC, shown by the dashed line.}
\label{fig:emergences}
\end{figure}

\section{Creating a catalogue of sunspots}
In order to analyse the sunspots over solar cycle 23, the STARA (Sunspot
Tracking and Recognition Algorithm) code developed by \cite{Watson2009} was
used, and readers are referred there for information on the method and its
testing. This is an automated system for detecting and tracking sunspots through
large datasets and also records physical parameters of the sunspots detected. It involves using techniques from the field of morphological image processing to detect the outer boundaries of sunspot penumbrae. This is achieved by means of the top-hat transform which allows us to remove any limb-darkening profile from the data and to perform the detections in one step. In
addition to the method given in \citet{Watson2009} the code had to be developed
further to separate the umbra and penumbra of spots as we would be looking at
the magnetic fields present in the umbra. When visually inspecting the
data there is a clear intensity difference between the umbra and
penumbra in sunspots. This difference
is due to the magnetic structure of susnpots. The umbra has a higher density of
magnetic flux which inhibits convection more than in the penumbral region. This
causes the umbra to be cooler and therefore appear darker. However, as sunspots
move towards the limb both the umbra and penumbra are limb-darkened. For this reason, we cannot use a single threshold value to define the
outer edge of the umbra. The algorithm we use removes all limb darkening effects at the same time as sunspot detection, greatly increasing speed as these two steps are carried out together. This problem has been approached by other authors
using different techniques, for example the inflection point
method of \cite{1997SoPh..171..303S}, the cumulative histogram method of
\cite{1997SoPh..175..197P}, the fuzzy logic approach of
\cite{2009SoPh..260...21F}, and the morphological approach of
\cite{2005SoPh..228..377Z}. Our method begins with the sunspots (which includes
umbrae and penumbrae) detected by STARA, and then produces a histogram of
sunspot pixel intensities for each spot. This clusters in two peaks, the local
minimum between which corresponds to the intensity value at the edge of the
umbra.  A similar histogram-based approach was implemented by
\cite{2009SoPh..260...21F} who then used concepts from fuzzy logic to assign
membership to umbra or penumbra; they showed that particularly the pixel
membership of the penumbra can vary significantly (tens of percent) depending on
a parameter known as the membership function, but this is apparently less of a
problem for low-resolution data, in which brightness variations within the
penumbra are smeared out. We have not adopted such a method, but have instead
identified the local minimum for each sunspot's histogram, and created a mask
for umbral pixels. We normally find that the umbra region of sunspots has an MDI pixel value of less than 7000 - 8000. However, our algorithm does have the benefit of being applied
consistently across the entire data series, and being able to deal with the
varying intensity across the solar disk due to limb darkening which eases the
problems of sunspot detection and area estimation that occur if a
straightforward intensity threshold is used.

The data used in this study are taken from the MDI instrument
\citep{Scherrer1995} on the SOHO spacecraft. We use the level 1.8 continuum data
as well as the level 1.8 magnetograms to analyse magnetic fields present in the
spots. Our dataset uses 15 years of data and we analyse daily measurements taken
at 0000UT when co-temporal continuum images and magnetograms are recorded. The
STARA code takes around 24 hours to process the approximately 5000 days of data
available to generate the sunspot catalogue used in this article and holds
30\,084 separate sunspot detections. The same sunspot will be detected in many
different images and tracked from image to image allowing them to be associated
with one another. The physical parameters obtained from this analysis are the
sunspot total area and `centre of mass' location, number and area of umbrae;
mean, maximum and minimum magnetic fields in the umbrae and penumbra; total and
excess flux in the umbrae and penumbra and the information relating to the
observation itself such as time and instrument used.

\subsection{Number of sunspots}
The trend of sunspot number throughout a solar cycle is well documented and
generally rises rapidly at the start of a solar cycle before a slower decrease
towards the end of the cycle. The Solar Influences Data Center (SIDC,
\url{http://www.sidc.be/sunspot-data/}) keeps records on the sunspot index and
so we compare
the results of our detections with the findings of the SIDC as an initial test.
It must be noted that both indicators are not measuring the same thing as the
international sunspot number recorded by the SIDC weights the sunspots seen in
groups so that it becomes a stronger proxy for solar activity whereas STARA only
gives us the raw number of observed sunspots. However, it is beneficial to see
if the same trends are present. The data used here are the smoothed monthly
sunspot number \citep{SIDC} and so our daily measurements have been treated in
the same way to give a fair comparison.

In Fig.~\ref{fig:emergences} we can see that both curves share several features.
The STARA output has been scaled up to the same level as the International
Sunspot Number around 2001 - 2003 when sunspot count rates were higher and the
general trends are more important here than absolute values due to the
differences in counting methods (this scaling is permissible due to the somewhat
arbitrary factors present in the SIDC sunspot numbers - see
Equation~\ref{eq:1}.) We see that both datasets exhibit the same patterns of
increasing and decreasing at the same time and the agreement is very good in the
declining phase of the cycle. This also continues into cycle 24 shown at the
right hand side of the plot with both curves rising at the same time and we will
continue to track the agreement of these further into the next cycle.

The SIDC data \citep{Clette2007}, shown as a dashed line on the plot has a
smooth rise up to the first maximum sometime in the year 2000 and falls before
reaching a second maximum in 2002. This `double maximum' feature, separated by
the `Gnevyshev gap' \citep{1967SoPh....1..107G} is also seen in the STARA output
although the first maximum is weaker when compared to the second, in contrast
with the SIDC data in which the first maximum is larger than the second.
However, both sets of data scale well with one another after this second maximum
with very little deviation and this continues from 2002 up to the current day.

The differences in the first peak, and indeed in the rise before that are most
likely due to the method of counting sunspots as mentioned previously. In fact,
the SIDC sunspot number is calculated using the formula

\begin{equation}\label{eq:1}
T = k ( 10g + s )
\end{equation}

where $T$ is the total sunspot number for that measurement, $g$ is the number of
sunspot groups observed and $s$ is the number of individual sunspots observed.
It is based on the assumption that sunspot groups have an average of 10 sunspots
in them and so even in poor observing conditions, this would be a good
substitute. The coefficient $k$ is a number that represents the seeing
conditions from the observing site and is usually less than 1.

What Fig.~\ref{fig:emergences} suggests is that the SIDC observers are either
detecting more sunspots than STARA in the first half of the cycle, or that they
are detecting groups that have fewer than 10 sunspots in them, on average. This
second explanation is more likely. Inspecting the STARA data we find it is rare
to see a sunspot group with as many as ten spots in this stage of the cycle,
which would account for the SIDC number being an overestimate for the actual
sunspot number at this time. This in itself has interesting implications for the
solar cycle, suggesting that very complex magnetic groups - and the heightened
activity that accompanies them - are more likely to appear in the second part of
the overall solar maximum.

\subsection{Sunspot locations}

\begin{figure}[ht]
\centerline{\includegraphics[width=0.45\textwidth,clip=]{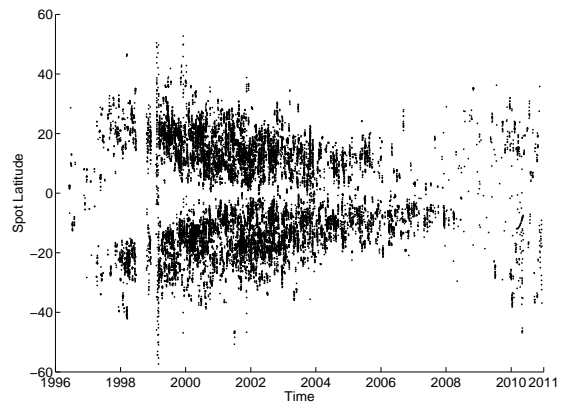}}
\caption{The latitude of all 30\,084 sunspot detections from solar cycle 23. The
end of solar cycle 22 can be seen as well as the onset of cycle 24. Note that
there is a much larger `gap' between cycle 23 and 24 than between cycles 22 and
23. This confirms the lack of solar activity from mid 2008 to early 2010.}
\label{fig:butterfly}
\end{figure}

The locations of sunspots were also recorded by the STARA code and this allows
us to produce a butterfly diagram of sunspot locations. The `butterfly' shape is
produced by the pattern of sunspot emergences seen in each cycle. At the start
of a cycle sunspots tend to appear at high latitudes, between 20 and 40 degrees
above and below the solar equator. But as the cycle progresses, the spot
emergences are observed closer to the equator. The cycle then ends before the
sunspots are seen to emerge at the equator and as a result of this it is very
rare to see a sunspot forming within a few degrees of the solar equator.
\citet{Zharkov2007} have observed a `standard' butterfly pattern in sunspot
emergences in cycle 23 and our results are shown in Fig.~\ref{fig:butterfly}

The butterfly shape can be clearly seen as can some other features. There are
gaps in 1998 as the SOHO spacecraft was lost for some time and no data were
recorded. Also, the vertical line in early 1999 corresponds to the failure of
the final gyroscope onboard and a rescue using gyroless control software. This
caused the spacecraft to roll and so all data recorded at this time does not
have a consistent sun orientation. These artifacts have been left in the figure
(although corrected for in our subsequent analysis) to illustrate some of the
potential problems with using long term data sets.

To enable the continuation of the mission the spacecraft is rotated
approximately every three months to allow the high gain antenna to point at the
Earth as it can no longer be moved. This means that the data are rotated and
this introduces further small errors in position detection as the roll angle is
not known exactly but the algorithm assumes that the data is either 'north up'
or 'south up'.

We can see from Fig.~\ref{fig:butterfly} that the end of solar cycle 23
exhibited asymmetric behaviour with very few spots appearing on the north
hemisphere compared to the south. \citet{Hathaway2010} shows that a north-south
asymmetry in sunspot area during a cycle is very common but he also states that
any systematic trend in the asymmetry during a solar cycle is found to change in
the next cycle and so is not particularly useful for predictions of activity or
for solar dynamo modelling. This asymmetry was studied in more detail by
\citet{Carbonell1993} using a variety of statistical methods and they found that
a random component was dominant in determining the trend of hemispheric
asymmetry in sunspots.

\subsection{Sunspot areas}

\begin{figure}[ht]
\centerline{\includegraphics[width=0.45\textwidth,clip=]{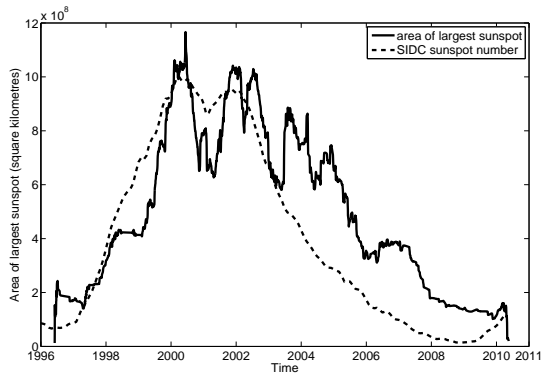}}
\caption{The area of the largest sunspot observed is shown here, smoothed over 3
months to minimise the effect of very large sunspots and days where no spots
were visible. This roughly follows the international sunspot number as well as
the activity seen throughout solar cycle 23.}
\label{fig:areas}
\end{figure}

As was the case with the number of sunspots detected, the area of the largest
visible sunspot also follows the activity of the solar cycle with a clear rising
phase and a slower declining phase. When calculating the area of a
sunspot or umbra the number of pixels within the spot or umbral boundary is
corrected to take into account the geometrical foreshortening effects that
change the observed area relative to its position on the solar disk.  We show
this in Fig.~\ref{fig:areas}. The variation is larger as sunspot sizes have a
larger range than the number of spots that are present. Again, this has been
smoothed to give a fair comparison to the international sunspot number
calculated by the SIDC. An interesting feature of this plot is that at the start
of cycle 24 there is no significant increase in the areas of observed spots so
we can say that there are more spots beginning to appear but the spot magnetic
fields are still weak.

\begin{figure}[ht]
 \centering
\begin{tabular}{ l
}\includegraphics[width=0.45\textwidth,clip=]{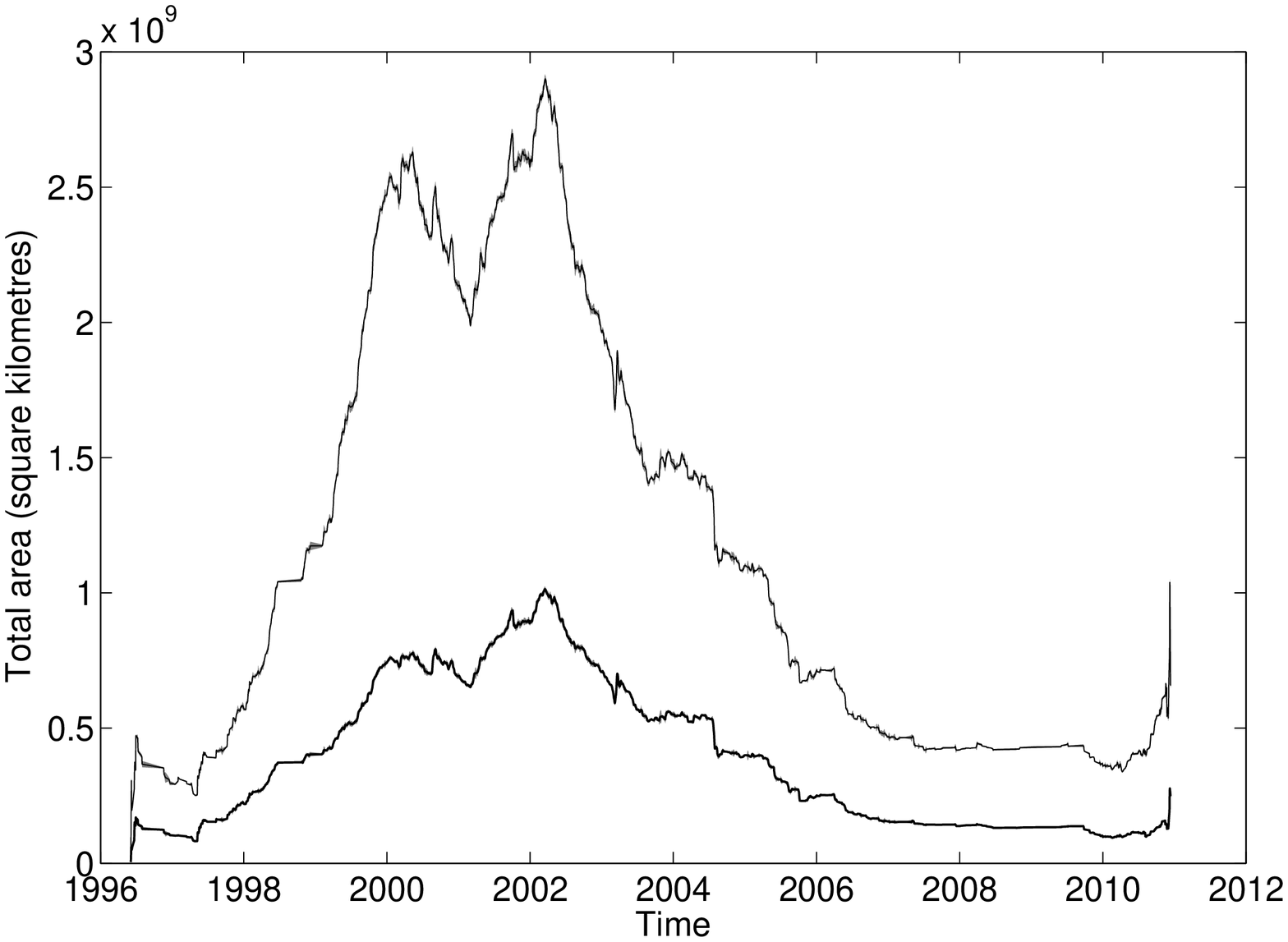}  \\
\includegraphics[width=0.45\textwidth,clip=]{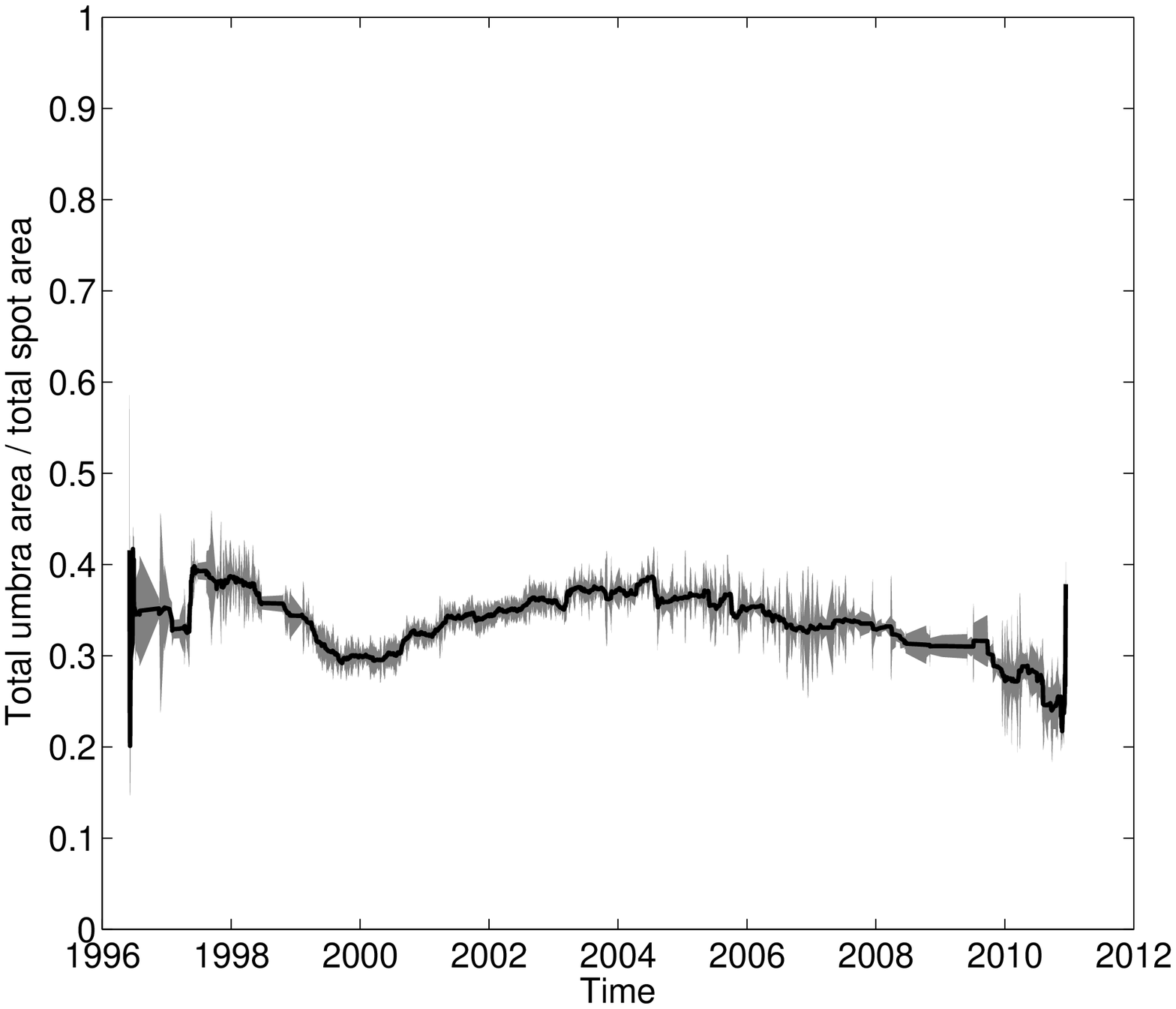}
\end{tabular}
\caption{top panel : The upper line shows the total observed sunspot area and
the lower line shows the total umbra area smoothed over three month periods and
corrected for foreshortening effects. Only sunspots within 60$^\circ$ of the
centre of the disk were used to minimise errors from this correction. bottom
panel : the ratio of total umbral area to total sunspot area. This ratio is
fairly constant, with the umbral area consiting of 30 - 40\% of the total
sunspot area and does not vary rapidly throughout the cycle. The errors are
shown by the shaded area and are lower between 1999 and 2005 due to the
increased number of sunspots at that time.}
\label{fig:ratio_areas}
\end{figure}

\begin{figure}[ht]
\centering
\includegraphics[width=0.45\textwidth]{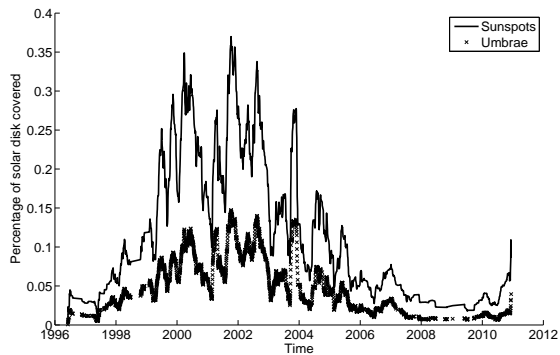}
\caption{We show the total sunspot (solid line) and umbra (crosses) area here as
a percentage of the area of the projected solar disk. The data are smoothed over
a three month period.}
\label{fig:deproj_areas}
\end{figure}

In addition to looking at the largest sunspot areas observed, we are also able
to examine the total area of the solar surface covered by sunspots at any one
time. This is shown in Fig.~\ref{fig:ratio_areas}. Both the total sunspot and
umbral areas are shown and, yet again, they both follow the overall trend of the
solar cycle with increases and decreases at the same times. More interesting
than this however, is the ratio of umbral area to sunspot area, shown in the
bottom panel. We observe that the umbral area is 20-40\% of the total observed
sunspot area and the ratio stays within this range throughout the cycle. Even
though a large variety of sunspot shapes and configurations are seen, the
fractional area of associated umbra does not show high amplitude fluctuations
unlike the maximum sunspot area observed - the dominant characteristic is a
relatively smooth variation. Note that this does not hold for individual
sunspots due to the variety of configurations seen, only to the large scale
distribution of sunspots over time. There are also interesting features present,
most of all the dip in the year 1999. At this time, the sunspot area is
increasing more quickly than the area of the associated umbrae. This soon
changes and the umbral areas start to occupy more of the sunspot again, rising
by a few percent by 2004 before starting to drop off again. During the first
peak in solar activity in 2000 we see that the umbra is occupying a lower
fraction of the sunspot and from Fig.~\ref{fig:emergences} this is when the
International Sunspot Number was higher than the STARA sunspot count. This could
indicate that there are sunspot groups with lower than ten sunspots present in
them. This suggests that there is more space in these groups for the sunspot
penumbrae to grow. In comparison to this, in the second peak of activity in 2002
we see that the fraction of sunspot area occupied by umbrae has grown and that
the STARA count rate is above the International Sunspot Number. This suggests
that we are seeing sunspot groups with more than ten spots in them. These would
be very complex groups and so it may be the case that the sunspots have multiple
umbrae present within them which would likely increase the fractional umbral
area.

In Fig.~\ref{fig:areas} and Fig.~\ref{fig:ratio_areas} we show the error in the
areas measured as a shaded band surrounding the line representing the data
points. Estimating the errors involved is done by examining the output of the
STARA algorithm. When detecting sunspots and sunspot umbrae, the centroid of the
region is determined with good accuracy. However, when defining the perimeter of
the region, we believe that there is an error of 1 pixel both towards and away
from the centre of the region. This means that large sunspots will have a
smaller fractional error than small spots, even though the absolute value of the
error will be greater for large spots. 

We also show the percentage of the projected solar disk covered by sunspots from
the viewpoint of the SOHO spacecraft in Fig.~\ref{fig:deproj_areas}. The trend
is very similar to that of the absolute total area of sunspots looked at
previously. We see the fraction of the solar disk covered by sunspots rise to
about 0.35\% at the peak of activity in cycle 23 which is equivalent to 3500 MSH
(millionths of a solar hemisphere). This is comparable to some of the largest
sunspots ever detected. There are significant short-term fluctuations in this
series, in addition to the overall solar cycle variation.

\section{The evolution of sunspot magnetic fields}
As the detection algorithm is directly linked with the MDI magnetograms recorded
at the same time, we are also able to track the evolution of the magnetic field
present in sunspots throughout the cycle. We assume that the magnetic field
within sunspot umbrae is in the local vertical direction. As the MDI data only
gives the line of sight magnetic field we apply a cosine correction to account
for this. The amplification of magnetic field strength due to the cosine
correction becomes very large as sunspots approach the limb, so making an
incorrect assumption about the field being vertical can lead to vastly wrong B
values at the limb. To minimise these effects we only include sunspots with a
value of $\mu > 0.95$ where $\mu$ is the cosine of the angle between the local
solar vertical and the observers line of sight. In addition to this, the
observed line of sight field is corrected with the assumption that the true
field direction is
perpendicular to the local photosphere. As we are looking at the strongest
fields in sunspot umbrae this is a reasonable approximation.

\begin{figure}[ht]
\centerline{\includegraphics[width=0.45\textwidth]{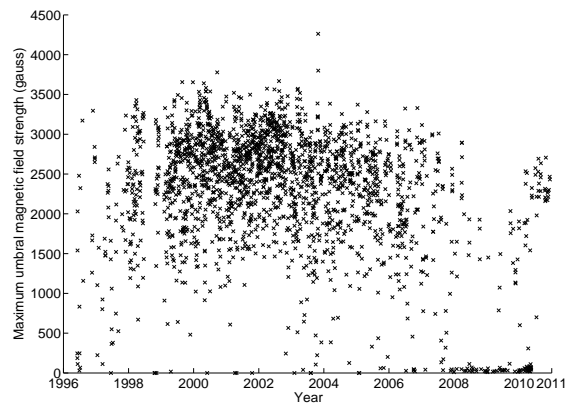}}
\caption{Maximum sunspot umbra field from 1996-2010. Measurements are taken
daily.}
\label{fig:rawspots}
\end{figure}

Fig.~\ref{fig:rawspots} shows the maximum sunspot umbral fields measured daily
from 1996 - 2010. The first thing to notice is the spread of magnetic fields
measured. We also see that the majority of measurements fall between 1500 and
3500 Gauss. It is very difficult to see any kind of trend in the data due to the
spread of values but we can observe a lack of strong sunspots from 2008 - 2010
when the most recent solar minimum occured.

\begin{figure}[ht]
\centering
\begin{tabular}{ l
}\includegraphics[width=0.45\textwidth]{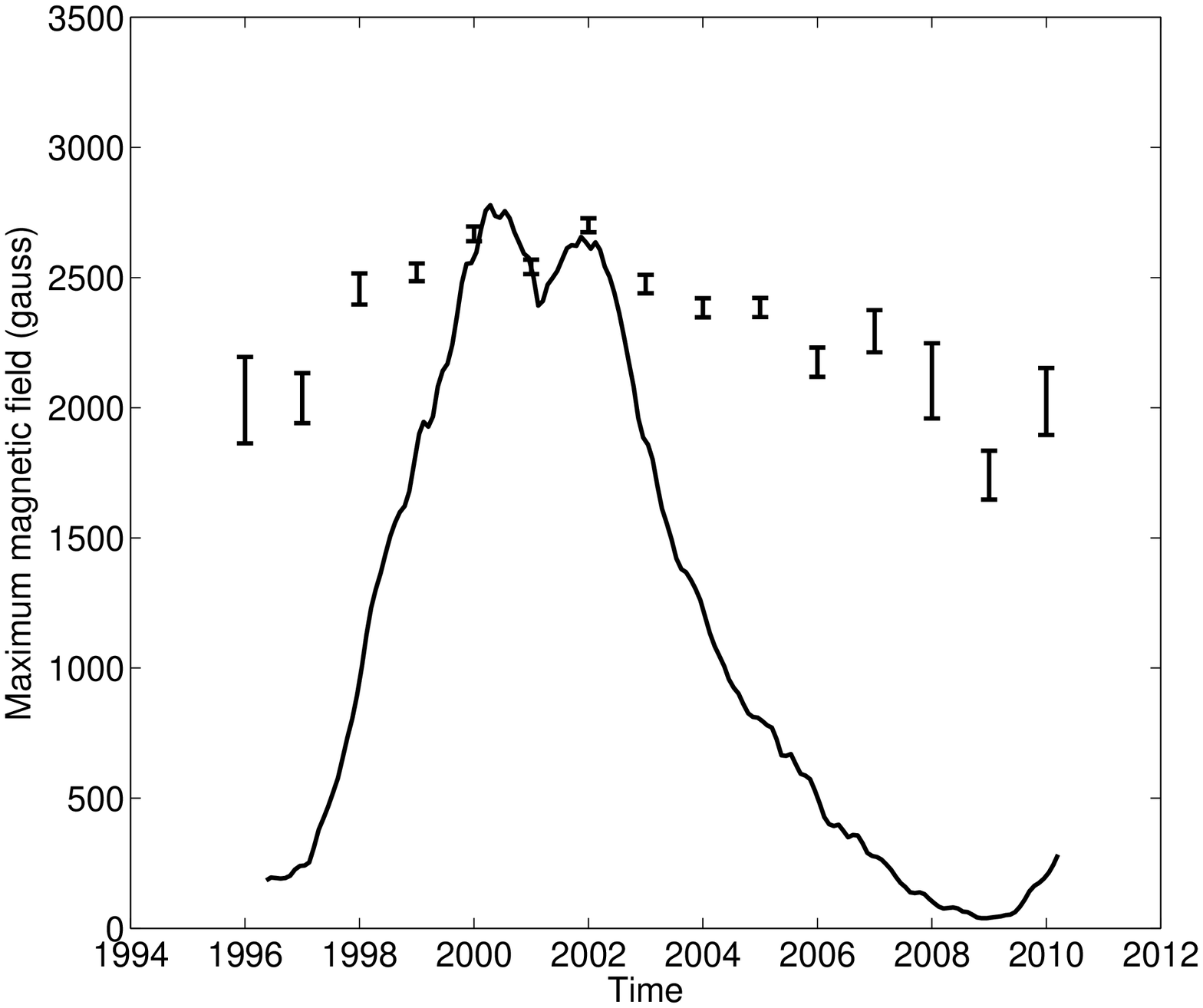} \\
\includegraphics[width=0.45\textwidth]{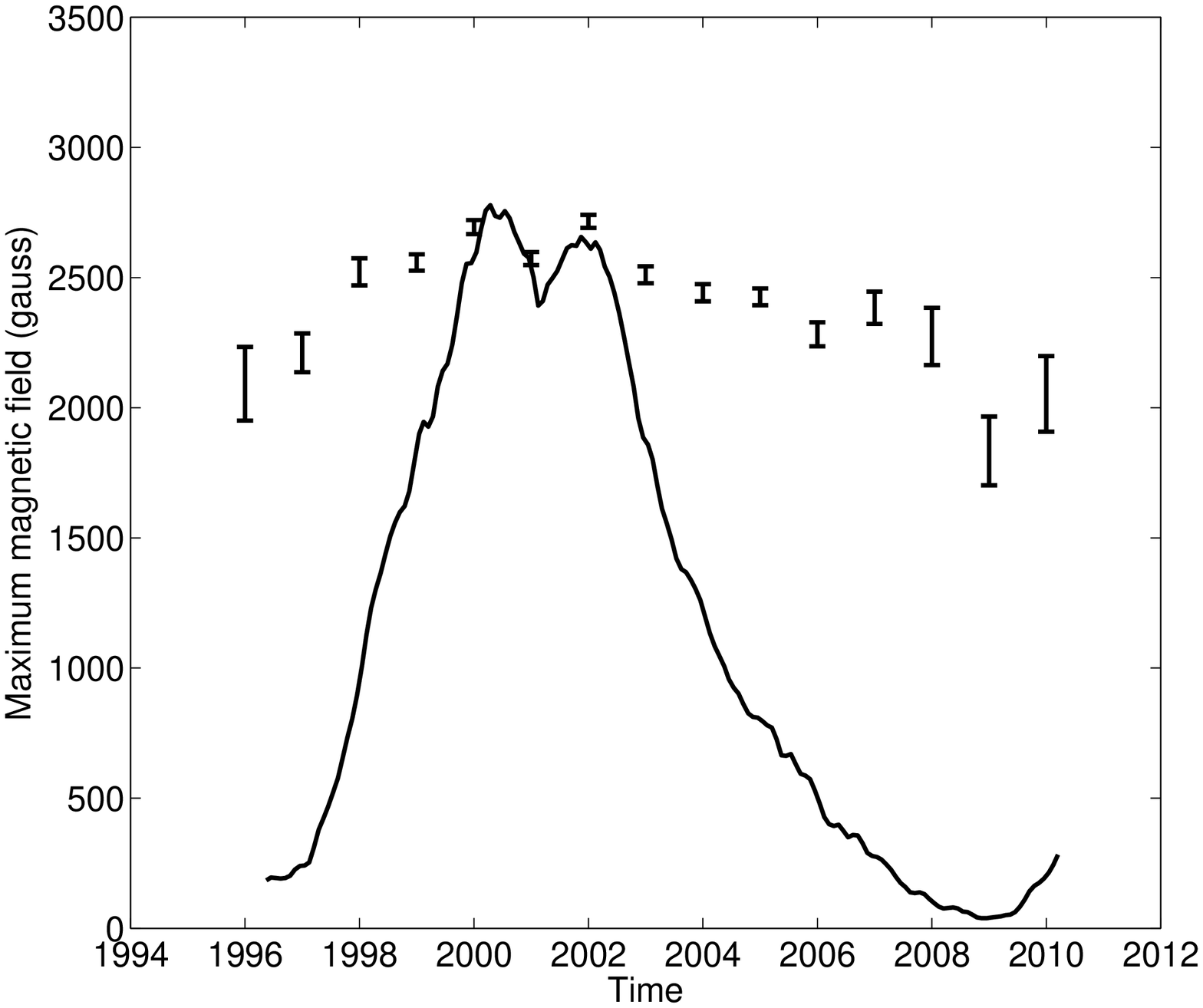}
\end{tabular}
\caption{The data shown in Fig.~\ref{fig:rawspots} have been binned by year and
the mean of each bin is plotted here. Top panel: all data from
Fig.~\ref{fig:rawspots} are included. Bottom panel: only measurements with a
field above 1500 Gauss are included. The error bars correspond to the standard
error on the mean. The solid line shows the evolution of the international
sunspot number over the same period for reference. Assuming a linear trend gives
a gradient of -23.6 $\pm $ 3.9 Gauss per year and -22.3 $\pm$ 3.9 Gauss per year
respectively.}
\label{fig:spottrend}
\end{figure}

A similar study has been undertaken by \citet{Penn2006} using the McMath-Pierce
telescope on Kitt Peak which includes umbra measurements going further back, to
1991. The method is different as they use the Zeeman splitting of the Fe I line
(1564.8nm) to infer a magnetic field strength at the location of the
measurement. Measurements are made in the darkest part of the umbra, where this
is identified in the image using a brightness meter. The Zeeman splitting
identified at that location is used to determine the true magnetic field as the
splitting of the spectral line observed is not dependent on the angle between
the magnetic field and the observers line of sight. Very small spots were
excluded from their dataset, as the small size of the umbra increases the risk
of scattering of penumbral radiation into the umbral area, and consequent
distortion of the line profile. Pore fields correspond to the range 1600-2600~G,
with a mean of 2100~G. When this dataset of maximum measured umbral field is
binned and averaged by year, and plotted as a function of time, a decrease is
visible which can be fitted with a linear trend equivalent to around -52 Gauss
per year. We repeat the analysis carried out by \cite{Penn2006} on our dataset,
both including and excluding all spots with a vertical magnetic field component
below 1500~G to minimise the possible effects of pores being included in the
analysis, for a direct comparison with the \cite{Penn2006} result. The results
are shown in Fig.~\ref{fig:spottrend}.

The top panel includes all sunspots detected whereas the bottom panel excludes
any spots with a maximum field strength of less than 1500 Gauss. The error bars
are calculated as the standard error on the mean of all measurements in the bin.

The data are in line with a picture in which the umbral fields are simply
following a cyclical variation pattern, as the increases and decreases follow
the international sunspot number. This cannot be confirmed with the current data
and we will need to wait until the next cycle is well under way to see if the
trends continues to be present. If we do a straight line fit as in
\citet{Penn2006}, then the gradient of the best fitting line gives a decrease in
umbral fields of 23.6 $\pm$ 3.9 Gauss per year which, although still decreasing,
is a far slower decline than seen by Penn and Livingston. Repeating the analysis
excluding sunspots with fields below 1500 Gauss gives a long term decrease in
field strength of 22.4 $\pm$ 3.9 Gauss per year. This is even further from the
result they observed, although as the sunspots with fields below 1500 Gauss make
up such a small fraction of the population we observe, we would not expect a
significant change in the result. Other studies have also cast doubt on the long
term decrease of umbral magnetic fields. The \citet{Penn2006} article suggests
that a decrease of 600 Gauss over a solar cycle would cause a change in mean
umbral radius as a relationship between these two quantities has been shown by
\citet{Kopp1992} and \citet{Schad2010} but follow up observations by
\citet{Penn2007} could not see this in their data. It has also been suggested by
\citet{Mathew2007} that a small sunspot sample may introduce a bias into results
if the size distribution of sunspots used is not calculated in advance.

However, the long term decline in sunspot magnetic fields does agree with the
lack of an increase in sunspot area as shown in Fig.~\ref{fig:areas}. If the
magnetic field is now weaker than at the same time in the last cycle we would
expect sunspots to be smaller and this is currently what is observed.

Interestingly, if the data from only the declining phase of the cycle (from 2000
to 2010) are used, then the maximum umbral field strengths are seen to decrease
by around 70 Gauss per year which is far greater than the \citet{Penn2006}
study.

This then leads to the question of how valid this comparison is. In
fact, instruments such as MDI and the new Helioseismic and Magnetic Imager on
SDO do not measure the true value of magnetic field strength in a pixel. The
value they return is an average magnetic field strength with a resolution
determined by pixel size. However, if the filling factor of spatially unresolved
magnetic elements within the pixel is close to unity, then the pixel value is a
good approximation for the true line of sight magnetic field strength. This is
thought to be the case deep in the umbrae of strong sunspots and so for these
measurements we can say that our observations are good approximations for the
true line of sight magnetic fields. In addition to this, we have only used
sunspots with $\mu > 0.95$ which corresponds to 18.2 degrees from solar disk
centre in an effort to minimise any corrections to the magnetic field
measurements but still assume that the field in the core of sunspot umbrae is
perpendicular to the local photosphere.

Also, MDI has problems with saturation in magnetic field measurements
with a peak value of between 3000 and 3500 Gauss depending on when the observation was made (the saturation value has lowered as the instrument degrades). This
has a greater effect on measurements made at solar maximum and so has the effect
of reducing the long term field strength decrease. However, this does not fully
account for the discrepancy between our value of the rate of long term field decrease and that of other studies.

We have not only compared the trends seen but also the data points used
in calculating these trends. The latest Livingston and Penn data is kept up to
date by Leif Svalgaard and can be viewed at his own website
(see \url{www.leif.org/research}). With the exception of a single data point in
1994, the Livingston and Penn yearly averages are similar to ours. Sadly, there are no
yearly averages in the Livingston and Penn data between 1994 and 2001 to better
compare the two studies.

\section{Discussion and Conclusions}
Using a catalogue of sunspot detections created by the STARA code provides a
reliable way to analyse the long term variation of certain physical parameters
relating to sunspots. We found that the number of sunspots detected compared
very well with the international sunspot number, even through the period of
2008-2010 when sunspot detections have been more sparse and difficult due to the
decreased magnetic field strengths that are causing them. When looking at the
locations of sunspots a traditional butterfly pattern is seen which also shows
the end of cycle 22 as well as the period of almost no sunspots from late 2008
to early 2010 before cycle 24 started. Fig.~\ref{fig:butterfly} also shows some
of the problems of a long term observing run, such as spikes in early 1999
caused by failure of the gyroscopes onboard SOHO. In addition to this, the high
gain antenna on SOHO malfunctioned in mid 2003. 

The area of sunspots was then examined with the maximum spot area being first
observed. The rough pattern of an initial steep rise and gradual fall associated
with a solar cycle was seen but with many other features present. However, when
the total observable sunspot area was plotted, a much smoother evolution was
seen. The same smooth evolution was also present in the total observable umbral
area. We also found that throughout the whole of solar cycle 23, if smoothed
over a three month period, the area of umbra visible was between 20 and 40\% of
the visible sunspot area once corrections for geometric foreshortening had been
applied.

We then continued to show the evolution of magnetic fields in sunspot umbrae and
Fig.~\ref{fig:rawspots} shows the large spread of sunspot magnetic fields
observed. Once the spot magnetic field data had been binned by year, a long term
cyclical trend could be observed but it is yet unknown whether this is a
cyclical variation around a long term linear decrease as suggested by other
studies. Our data supports
stronger fields near solar maximum and weaker fields at solar minimum. When
compared with other similar studies, the rate of magnetic field decrease is very
different and is likely due to the wide range of sunspot fields. The next solar
cycle should bring a more definitive answer to the question of whether a secular
trend in sunspot fields exists over multiple solar cycles. We will continue to
track this for as long as SOHO still flies and also plan to incorporate data
from the new Helioseismic and Magnetic Imager on the Solar Dynamics Observatory
spacecraft which serves as the successor to SOHO.

\begin{acknowledgements}
F.T.W. acknowledges the support of an STFC Ph. D. studentship. This work was
supported by the European Commission through the SOLAIRE Network
(MRTN-CT-2006-035484) and by STFC rolling grant STFC/F002941/1. SOHO is a
project of international cooperation between ESA and NASA. We acknowledge
useful discussions with M. Hendry and would like to thank our anonymous referee
for their thought provoking comments. Thanks also to Leif Svalgaard for allowing
us
to use his plots for comparing with Livingston and Penn data.
\end{acknowledgements}

\bibliographystyle{aa}
\bibliography{SunspotCataloguePaperRefs}
%\bibliography{sunspot_paper}
\end{document}